 \documentclass[final,1p,times,authoryear]{elsarticle}

\usepackage{amssymb}
\usepackage{color}

 \usepackage{lineno}

\journal{Ocean Modelling}

\def\upi{\pi}
\def\dd{{\mathrm d}}    
\def\rv{{\mathbf r}}   
\def\kv{{\mathbf k}}  

\begin{document}

\begin{frontmatter}



\title{The generalized Phillips' spectra and new dissipation function for wind-driven seas}

 \author[aff1,aff2,aff3,aff4,aff5]{Vladimir E. Zakharov}
 \author[aff1,aff3,aff4]{Sergei I. Badulin\corref{cor1}}
 \address[aff1]{P.P.Shirshov Institute of Oceanology of Russian
Academy of Sciences}
 \address[aff2]{University of Arizona, Tuscon, AZ, USA}
\address[aff3]{Laboratory of Nonlinear Wave Processes, Novosibirsk State University, Russia}
\address[aff4]{P.N. Lebedev Physical Institute of Russian
Academy of Sciences}
\address[aff5]{Waves and Solitons LLC, 1719 W.Marlette Ave., Phoenix, AZ 85015, USA}

\cortext[cor1]{Corresponding author Sergei I. Badulin}
\ead{badulin@ioran.ru}

\begin{abstract}
A generalization of the kinetic equation   is proposed for explaining observed shapes of wind wave spectra. The approach allows to fix a critical uncertainty in modeling wind wave spectra using a  condition of equilibrium of nonlinear transfer and wave dissipation due to breaking. We demonstrate transition from the Kolmogorov-Zakharov spectrum $E(\omega)\sim \omega^{-4}$ to the Phillips spectrum $E(\omega)\sim \omega^{-4}$. This transition is routinely observed in field experiments. The first results of the generalized kinetic  equation simulations are presented.
\end{abstract}


\end{frontmatter}




\section{Introduction}
\label{sect:Intro}
Wind waves in seas and lakes represent a complicated natural phenomenon governed by a great number of physical mechanisms. Complete physical theories for many of these mechanisms are not proposed yet that makes particular asymptotic cases extremely valuable.  \cite{Phillips58} considered a limiting case of saturation of a random wave field due to wave breaking. The dimensional consideration gave the famous Phillips energy spectrum $E(\omega) =\alpha g^2 \omega^{-5}$. Phillips himself emphasized the idea of `hard, saturated limit of the spectrum' where `any excursion above this limit' is `relieved immediately by breaking' \cite[see p.506 of ][]{Phillips85}.

The conceptual passage from the idea of saturation to  the concept of an equilibrium of random wave field brought \cite{Phillips85} himself to the model of spectral balance where all the mechanisms, wave input by wind, dissipation due to breaking and nonlinear transfer due to resonant wave-wave interactions compete on equal terms. In his sophisticated and somewhat eclectic model \cite{Phillips85} tried to explain other fundamental exponent $-4$ of wind-wave spectra found in previous theoretical \citep{Kitai62,Toba1973a}  and experimental \citep{Kawai1979,Mitsuyasu80,Forristall81,Kahma81} studies. The essential flaw of the Phillips theoretical construction of spectral equilibrium, in its own words \cite[sect.5 of ][]{Phillips85}, is a number of  empirical parameters responsible for wind-wave spectral balance. All these parameters should be matched each other. Phillips formulated the corresponding criteria of consistency of his model but pointed out the difficulties of this matching and deficiency of our knowledge of wind-wave physics. The validation of this well-known and widely accepted model remains a big problem and is not completed so far. Due to this fact, in particular,  the model cannot be extended to the case of the classic Phillips spectrum $\omega^{-5}$ in a consistent manner. Thus, one has a conceptual gap between the quasi-linear Phillips' vision of spectra  close to $\omega^{-4}$ and the `utterly nonlinear' authentic \cite{Phillips58} spectrum $\omega^{-5}$. The present paper is aimed at filling this gap following the theory of wave turbulence \citep{ZakhFalLvov92} and its recent extension \citep{NewellZakharov2008}.

In his thorough and prophetic study Phillips missed a key physical feature of nonlinear transfer in wind-wave spectra: the wave-wave interactions are able to provide an equilibrium state `on their own' in absence of wind input or wave dissipation. Thus, the  weakly turbulent Kolmogorov-Zakharov cascade solutions \citep{ZakhFil66,ZakhZasl83a} appear to be `self-sufficient' for explaining exponents $-4$ of direct  and $-11/3$ of inverse cascades as inherent features of deep water wave spectra governed by constant spectral fluxes in an equilibrium range of wave scales. Thorough analysis of wind-wave balance shows that this equilibrium range where nonlinear resonant interactions are dominating does exist for typical conditions of wind seas and for wave scales we are really interested in \citep{YoungVledder93,BPRZ2005}. The corresponding arguments have been presented recently both in terms of physical scales of nonlinear transfer \citep{ZakharovBadulin2011DAN} and universal links of wind wave parameters \citep{Badulin2014Univ}.

A remarkable fact is that the weakly turbulent approach can be extended beyond the equilibrium range to the domain where an alternative mechanism competes with the nonlinear transfer  on equal terms. This concept of the generalized Phillips' spectra has been developed recently by \cite{NewellZakharov2008}. The basic assumptions of the weak turbulence theory  can keep their formal validity for infinitely long time in the whole range of wave scales. These special solutions provide an alternative treatment of the classic Phillips spectrum $\omega^{-5}$ within the weak turbulence theory rather than within the concept of water surface singularities due to breaking.

In this paper we use the extension of the wave turbulence theory  by \cite{NewellZakharov2008} for explaining the authentic Phillips' spectrum $\omega^{-5}$ as a result of competition of wave-wave resonant interactions and inherently nonlinear dissipative processes. The new dissipation function can be found in physically consistent way as one corresponding to the generalized Phillips' spectrum.  This new dissipation function has remarkable features. First, the balance of wave-wave resonances and nonlinear dissipation fixes the dissipation rate for the classic Phillips spectrum $\omega^{-5}$. This surprising result is in dramatic contrast with conventional  Phillips model where the dissipation rate is assumed to be arbitrary. Secondly, within the new model of this paper the dissipation absorbs completely the spectral energy flux directed to infinitely high wave numbers. Thus, the extended wave turbulence approach appears to be able to describe the whole range of wave scales and explain different exponents of the observed wave spectra \citep{Liu1988,Liu1989}.

In \S~2 we give  basic equations and necessary comments to the concept of the generalized Phillips' spectra. A simple isotropic model of the generalized Phillips spectrum of deep water waves is discussed in \S~3. Stationary and self-similar solutions for this model are analyzed as a basis for further comparison with results of simulations. Results of the simulations of wind-wave spectra within the kinetic equation  with the new dissipation function are presented in \S~4.  Discussion and conclusions are given in \S~5.

\section{Generalized Phillips' spectrum of deep water waves}
In this section we reproduce results of \cite{NewellZakharov2008} for a problem of statistical description of wind-wave field.  We consider the Hasselmann kinetic equation for spatial spectrum of wave action $N_\kv$ of wind-driven waves in the following form
\begin{equation}
\label{eq:Kinfull} \frac{\partial N_{\kv}}{\partial t} +
\nabla_{\kv} \omega_{\kv} {\nabla_\rv N_{\kv}} = S_{nl}+S_{in}+S_{diss}
\end{equation}
Subscripts $\kv,\,\rv$ for  $\nabla$ are used for gradients in wavevector $\kv$  and in coordinate $\rv$  correspondingly. For $N_\kv(t)$ and $\omega_\kv$ the subscript $\kv$ means dependence on wavevector. The term $S_{nl}$ in (\ref{eq:Kinfull})  is responsible for four-wave resonant interactions. Terms $S_{in}$ and $S_{diss}$ represent correspondingly input of wave action from wind and its dissipation. The description of $S_{in}$ and $S_{diss}$ is based on phenomenological parameterizations mainly \cite[see][]{WisePaper2007}. It gives very high dispersion of estimates of the terms \cite[see fig.~1 in][]{BPRZ2005} and a great number of physically different compositions of  $S_{nl},\, S_{in},\, S_{diss}$. Validity and physical correctness of the empirically based terms $S_{in}$ and $S_{diss}$ are generally beyond critical consideration: quantitative aspects are dominating over obvious questions on their physical relevance. It is not the case of the theoretically based term $S_{nl}$ derived from the first principles for assumptions formulated explicitly. In many cases validity of these assumptions can be checked directly, say, in simulations \citep{KorotkZakh2013}.

For deep water waves four-wave resonant interactions are described by the collision integral
\begin{equation}\label{eq:coll_int}
\begin{array}{lll}
    S_{nl}&=&\upi g^2 \int |T_{0123}|^2\left(N_0 N_1 N_2+N_1 N_2 N_3-N_0 N_1 N_2 - N_0 N_1 N_3\right)\\[7pt]
    &\times& \delta(\kv + \kv_1 - \kv_2 - \kv_3)\delta(\omega_0+\omega_1-\omega_2 -\omega_3)d \kv_1 d\kv_2 d\kv_3
    \end{array}
\end{equation}
Here we use convention of \cite{Zakharov2010Scr} for normalization of spectral densities $N(\kv)$ and kernels  $T_{0123}$. We also use subscripts $i=0\ldots 3$ to denote dependence on argument $\kv_i$ \cite[see][]{Krasitskii1994}.
Collection of   cumbersome expressions for the interaction kernel $T_{0123}$ can be found in Appendix of \cite{BPRZ2005}.

\subsection{Homogeneity property of $S_{nl}$ and validity of asymptotic statistical description}
Homogeneity properties of $S_{nl}$ are of key importance for our consideration. First, wave frequencies and wavevectors in (\ref{eq:coll_int}) obey the power-law dispersion relation for linear deep water waves
\begin{equation}\label{eq:disp_ww}
    \omega(\kv)=\sqrt{g |\kv|}.
\end{equation}
Interaction kernel $|T_{0123}|^2$ is also a homogeneous function of power $6$, i.e.
\begin{equation}\label{eq:kernel_hom}
    |T(\kappa \kv_0,\kappa \kv_1,\kappa \kv_2,\kappa \kv_3)|^2=\kappa^6 |T( \kv_0, \kv_1,\kv_2, \kv_3)|^2
\end{equation}
Correspondingly, for the collision integral (\ref{eq:coll_int}) that is cubic in wave action spectral density one has the well-known re-scaling property in terms of wavenumber dependence
\begin{equation}\label{eq:col_hom_k}
    S_{nl}[\kappa \kv, \nu N_\kv] = \kappa^{19/2} \nu^3 S_{nl}[ \kv,  N_\kv].
\end{equation}
or in terms of dependence on frequency $\omega$
\begin{equation}\label{eq:col_hom_w}
    S_{nl}[\upsilon \omega, \nu N_\omega] = \upsilon^{11} \nu^3 S_{nl}[ \omega,  N_\omega]
\end{equation}
One can see that the collision integral depends heavily on wave scales and amplitudes that may break validity of the asymptotic approach during long-time evolution of wave field.  One of the basic assumptions of the asymptotic approach is smallness of wave period $T$ as compared with time scale of nonlinear interactions, i.e.
\begin{equation}\label{eq:validity_ke}
    \frac{T}{T_{nl}}=\frac{1}{\omega_\kv N_\kv} \frac{dN_\kv}{d t}=\frac{1}{\omega_\kv N_\kv} S_{nl} \ll 1
\end{equation}
It can be checked  that for
\refstepcounter{equation}
$$
N_\kv \sim |\kv|^{-9/2} \quad \textrm{or} \quad N_\omega \sim \omega^{-5}
\eqno{(\theequation{\mathit{a},\mathit{b}})}
\label{eq:Phillips_spec}
$$
conditions (\ref{eq:col_hom_k},\ref{eq:col_hom_w}) are satisfied for any stretching parameter $\kappa$ and $\upsilon$ in (\ref{eq:col_hom_k},\ref{eq:col_hom_w}). In other words, the asymptotic approach appears to be formally valid at any wave scale. Moreover, one can prove that this is true at any order of the asymptotic approach. For the collision integral asymptotic series
\begin{equation}\label{eq:Snl_asym}
    S_{nl}=\sum_{n=4}^{\infty} S_{nl}^{(n)}
\end{equation}
conditions similar to (\ref{eq:validity_ke}) remain valid for every term $S_{nl}^{(n)}$ that represents resonant interaction of $n$ waves  \citep{NewellZakharov2008}. Thus, the generalized Phillips spectrum (\ref{eq:Phillips_spec}) correspond to a special case when asymptotic description of wave-wave interaction can be formally extended over the whole range of wave scales. It gives a chance to propose a model where weak nonlinearity acts as a strong physical factor and is able to concur with other strong physical mechanisms, say, with dissipation due to wave breaking.

It should be stressed that the found Phillips spectrum for deep water waves differs from the classic Kolmogorov-Zakharov solutions for direct and inverse cascading \citep{ZakhFil66,ZakhZasl83a}
\begin{eqnarray}
& N^{(1)}(\kv) &= C_p {P^{1/3}}{g^{-2/3}}{|\kv|^{-4}} \label{eq:Kolmogorov_direct}\\[7pt]
& N^{(2)}(\kv) &= C_q {Q^{1/3}}{g^{-1/2}}{|\kv|^{-23/6}}
\label{eq:Kolmogorov_inverse}
\end{eqnarray}
($P,\,Q$ -- energy and wave action  flux, $C_p,\,C_q$ -- the corresponding Kolmogorov's constants).
Collision integral $S_{nl}$ for  solutions (\ref{eq:Kolmogorov_direct}, \ref{eq:Kolmogorov_inverse}) is plain zero and estimates of the kinetic equation validity (\ref{eq:validity_ke}) requires special care. Following the simplest (but not trivial) way proposed recently in \cite{Zakharov2010Scr,ZakharovBadulin2011DAN} one can split  the nonlinear transfer  term $S_{nl}$ into two parts -- nonlinear forcing $F_\kv$ and definitely positive term of nonlinear damping $\Gamma_\kv N_\kv$ ($\Gamma_\kv$ -- nonlinear damping rate) as follows
\[
S_{nl}=F_{\kv} - \Gamma_\kv N_\kv.
\]
Evidently, the relaxation rate $\Gamma_\kv$ gives a physically correct estimate of time scale of nonlinear wave-wave interactions in the kinetic equation (\ref{eq:Kinfull}). Thus, the KZ solution  breaks when \cite[see eq.17 in ][]{ZakharovBadulin2011DAN}
\[
\Gamma_\kv \omega \simeq 4\upi g |\kv|^{9} N_{\kv}^2 \simeq 1
\]
that gives
\[
4\upi C_p^{2} g^{-1/3}P^{2/3} |\kv_{br}| \simeq 1
\]
for direct cascade solution  by \cite{ZakhFil66} and
\[
4\upi C_q Q^{2/3} |\kv_{br}|^{4/3} \simeq 1
\]
for the inverse cascade  by \cite{ZakhZasl83a}. The generalized Phillips' spectrum (\ref{eq:Phillips_spec}) being steeper than its KZ counterparts cannot be balanced `on its own'. In contrast to  KZ solutions (\ref{eq:Kolmogorov_direct}, \ref{eq:Kolmogorov_inverse}) it requires an additional term to provide a full balance in the right-hand side of the kinetic equation.

\subsection{Matching the dissipation function with nonlinear transfer term}
Matching a dissipation function can be done from different physical positions. First, one can rely completely on dimensional consideration for constructing a nonlinear dissipation function in the spirit of conventional wind-wave models as a local nonlinear function of spectral density $N(\kv)\, (N(\omega))$, frequency $\omega $ (or wavenumber $|\kv|$ and gravity acceleration $g$ as the only physical scale of deep water waves. To match homogeneity conditions (\ref{eq:validity_ke}) of the nonlinear transfer term for the (\ref{eq:Phillips_spec}) spectrum  one has
\begin{eqnarray}\label{eq:diss_PhillipsNk}
    S_{diss}[\kv,N_\rv] &= & \gamma_\kv N_\kv = \Phi_\kv (g^{1/2}{\kv}^{9/2} N_\kv) \omega N_\kv, \\[5pt]
        \label{eq:diss_PhillipsNw}
    S_{diss}[\omega,N_\omega] & = & \gamma_\omega N_\omega = \Phi_\omega({\omega}^{6} N_\omega/g^2) \omega N_\omega
\end{eqnarray}
with  arbitrary functions $\Phi_\kv$ or $\Phi_\omega$ of non-dimensional arguments.  Such `conventional' approach implies that wave dissipation is determined by instantaneous state of wave field and, in a sense, represents the \cite{Phillips58} authentic idea of saturated spectra.

An alternative way to introduce the nonlinear dissipation is to follow the Phillips idea of an equilibrium of different physical mechanisms, in our case, wave dissipation and nonlinear transfer. It can be regarded as `more physical': we assume that flux of energy (or wave action) due to nonlinear transfer is damped by nonlinear dissipation. Thus, one has a dynamical equilibrium provided by these two physical mechanisms. Such vision of wave balance can be formulated quite naturally in terms of wave turbulence where spectral flux of energy $P$ plays a key role
\begin{equation}\label{eq:New_diss}
    S_{diss}[\omega,E_\omega]= \Psi({P \omega^3}/{g^2}) \, \frac{P}{\omega}
\end{equation}
where $\Psi$ is arbitrary function of non-dimensional argument based on the flux $P$.

\section{The Phillips dissipation function and Phillips' spectrum }
Consider the simplest model of balance of nonlinear transfer and wave dissipation (\ref{eq:New_diss}) that satisfy condition (\ref{eq:validity_ke}) and provides the Phillips spectrum (\ref{eq:Phillips_spec}) formally in the whole range of wave scales
\begin{equation}
\label{eq:Kinext}
\frac{\dd E(\omega)}{\dd t}=-\frac{\partial P(\omega)}{\partial \omega} - \Psi(P\omega^3/g^2)\frac{ P(\omega)}{\omega}
\end{equation}
The right-hand side of (\ref{eq:Kinext}) operates with formally non-local quantity of spectral flux of wave energy
\begin{equation}\label{eq:def_energy_flux}
    P(\omega)=- \int_0^\omega \int_{-\pi}^{\pi} S_{nl}[\omega,\theta,E_\omega] d\omega d\theta
\end{equation}
In fact, this non-local value can be interpreted quite naturally using homogeneity properties (\ref{eq:col_hom_k}, \ref{eq:col_hom_w}). For gravity waves $ P\sim E^3(\omega) \omega^{12}$ (see \ref{eq:Kolmogorov_direct}) one has
\begin{equation}\label{eq:P2mu}
    \frac{P\omega^3}{g^2} \sim \frac{E^3(\omega)\omega^{15}}{g^6}=\mu_w^6
\end{equation}
i.e. the non-dimensional argument of $\Psi$ is proportional to wave steepness in power $6$. Note, that we use differential frequency-dependent steepness $\mu_w$ in (\ref{eq:P2mu}) in contrast to integral steepness
\begin{equation}\label{eq:mup}
    \mu_p^2=\frac{E_{tot}\omega_p^4}{g^2}
\end{equation}
($\omega_p$ spectral peak frequency) or cumulative steepness
\begin{equation}\label{eq:muc}
    \mu_c^2=g^{-2}\int_0^\omega E(\omega) \omega^4 \dd \omega.
\end{equation}
The steepness definition (\ref{eq:mup}) in terms of total energy $E_{tot}$ gives always finite values. Typically, for growing wind sea it is below $0.1$ and is decaying with time or fetch. Two alternative definitions (\ref{eq:P2mu},\ref{eq:muc}) are useful for description high-frequency part of wave spectra. For weakly turbulent direct cascade solution (\ref{eq:Kolmogorov_direct}) the differential and cumulative  steepnesses are infinitely growing functions $\mu_w \sim \mu_c \sim \sqrt{\omega}$. The Phillips spectrum $E(\omega)\sim \omega^{-5}$ gives saturation of differential steepness $\mu_w$ and logarithmic growth of $\mu_c$.

Our interpretation of function $\Psi(P\omega^3/g^2)$  equalizes two different models of wave dissipation: local in spectral density (\ref{eq:diss_PhillipsNk}, \ref{eq:diss_PhillipsNw}) and non-local flux-dependent one (\ref{eq:New_diss}). We fix this parallel below in the notation of wave steepness
\[
\mu_w=\frac{P\omega^3}{g^2}=\mu.
\]

\subsection{Stationary solutions of the generalized kinetic equation}
The stationary solution of (\ref{eq:Kinext}) for an arbitrary dependence $\Psi(\mu)$ gives nothing but dependence of steepness $\mu=P\omega^3/g^2$ on wave frequency
\begin{equation}\label{eq:stat_sol_gen}
    \int_0^{\mu}\frac{6\dd \mu}{\mu\left(3-\Psi(\mu^6)\right)}=\ln\left(\frac{\omega}{\omega_1}\right).
\end{equation}
A finite $\omega_1 $ in (\ref{eq:stat_sol_gen}) is introduced from dimensional consideration only. The denominator of the integrand in (\ref{eq:stat_sol_gen}) must not be zero at finite frequency $\omega$ for the integrand convergence and, hence, for physically correct values of $\mu$ (strictly positive). Thus, function $\Psi(\mu)$ has to be tending to a constant value. Any cases but $\Psi(\mu)\to 3$ give power-law dependence of $\mu$ on frequency $\omega$. As far as $\mu$ has a sense of wave steepness the case $\Psi(\mu)\to 3$ is seen as the only physically relevant one. In this case only the steepness for infinitely short waves is decaying with $\omega$.

Important results can be obtained for power-law dependence of function $\Psi$ on non-dimensional spectral flux $P\omega^3/g^2$. Let
\begin{equation}\label{eq:Psi_power}
    \Psi=C\left(\frac{P\omega^3}{g^2}\right)^R
\end{equation}
Stationary solution of (\ref{eq:Kinext}) obeys
\begin{equation}\label{eq:stat_sol_pow}
    C\left(\frac{P\omega^3}{g^2}\right)^R=3
\end{equation}
Thus, the dissipation function $\Psi(P\omega^3/g^2)$ should be constant for the stationary solution for any exponent $R$ in (\ref{eq:Psi_power}).  Note, that for $R=0$ in (\ref{eq:Psi_power})  the formal stationary solution gives an arbitrary exponent that is determined by coefficient $C$, i.e.
\[
P\sim \omega^{-C}.
\]
Mathematically, the arbitrary coefficient $C$ can give an arbitrary  exponent for frequency spectrum. But from physical viewpoint we should take care because at $R=0$ the key physical parameter -- gravity acceleration $g$ falls out of the model. To avoid `the non-physical freedom' one should take a formal limit at $R \to 0$ that fixes the value  $C=3$.

For $R \ne 0$ exponent $(-3)$ and dissipation rate $C$ are fixed by condition (\ref{eq:stat_sol_pow}). Surprisingly, the condition of stationarity does not depend on choice of power-like function $\Psi$, i.e. on dependence of dissipation on wave steepness.

From homogeneity property (\ref{eq:P2mu}) one gets immediately the Phillips spectrum
\begin{equation}
\label{eq:Phil_spec}
E(\omega) = \alpha_{Ph}g^2 \omega^{-5}
\end{equation}
where the Phillips $\alpha_{Ph}$ can be determined from the  relationship between energy spectrum and spectral flux.

\subsection{Self-similar solutions for the generalized kinetic equation}
Stationary solutions cannot be considered as a final argument for the Phillips spectrum: solutions that evolve in time and space should be studied as routes to possible equilibria  of wave spectra. Fortunately, for $R=0$ the generalized kinetic equation (\ref{eq:Kinext}) has self-similar solutions and re-scaling property (\ref{eq:P2mu})  can be used in its full.

Consider spatially homogeneous (the so-called duration-limited) solutions of (\ref{eq:Kinext}) in the form
\begin{equation}\label{eq:ss_form}
E \sim t^{p+q} U_p(\omega t^q)
\end{equation}
Total energy of this solution grows as
\begin{equation}\label{eq:Etot_growth}
   E_{tot} \sim t^p
\end{equation}
Looking for  homogeneity properties  of (\ref{eq:Kinfull}) one gets from (\ref{eq:P2mu}) the following conditions for exponents $p,\,q$
\begin{equation}\label{eq:ss_1}
    p=\frac{9 q-1}{2}
\end{equation}
that is the same as for the family of self-similar solutions of the authentic conservative kinetic equation \cite[see][for details]{BPRZ2005}. The second condition single out special exponents for self-similar solutions of the generalized kinetic equation (\ref{eq:Kinext})
\begin{equation}\label{eq:ss_2}
R(q-1)=0
\end{equation}
The exponent $q=1$ is unrealistic as soon as it corresponds to very fast growth of energy ($E_{tot}\sim t^4$) while $R=0$ looks quite attractive for the model of wave dissipation.

The advantage of the proposed dissipation function in (\ref{eq:Kinext})
\begin{equation}\label{eq:Sss}
    S_{diss}=-C_{Ph} \frac{P(\omega)}{\omega}
\end{equation}
is two-fold. First, self-similar solutions are seen usually as  physically relevant asymptotics of solutions with arbitrary initial and boundary conditions. The second feature of the dissipation function is more important for further study. While the nonlinear transfer and dissipation terms have similar properties of homogeneity they can be balanced without additional fitting parameters. Thus, choosing the dissipation function (\ref{eq:Sss}) we follow the \cite{Phillips85} idea `like cures like'.

Note, that \cite{Phillips85} realized this principle for all three terms in the right-hand side of the kinetic equation by fitting essentially nonlinear collision integral $S_{nl}$ and quasi-linear functions  $S_{in},\,S_{diss}$ to each other. The outcome of such composition was the well-known energy spectrum $E(\omega)\sim \omega^{-4}$ associated with a number of experimental findings \cite[e.g.][]{Toba1973a,DonelanHui85,BattjesEtal1987}. Extremely high price for this sophisticated composition  was a number of uncertainties in empirical parameters of functions of wave input and dissipation \cite[see sect.5 `Constraints on the empirical constants' in][]{Phillips85}.

Following the Phillips principle `like cures like' for the case of self-similar evolution with dissipation function (\ref{eq:Sss}) we avoid the Phillips problem:
\begin{center}
\emph{the spectral balance in our model does not require any fitting parameters}.
\end{center}

\subsection{Universality of dissipation function for the Phillips spectrum $E\sim \omega^{-5}$}
\begin{figure}
    \includegraphics[scale=0.5]{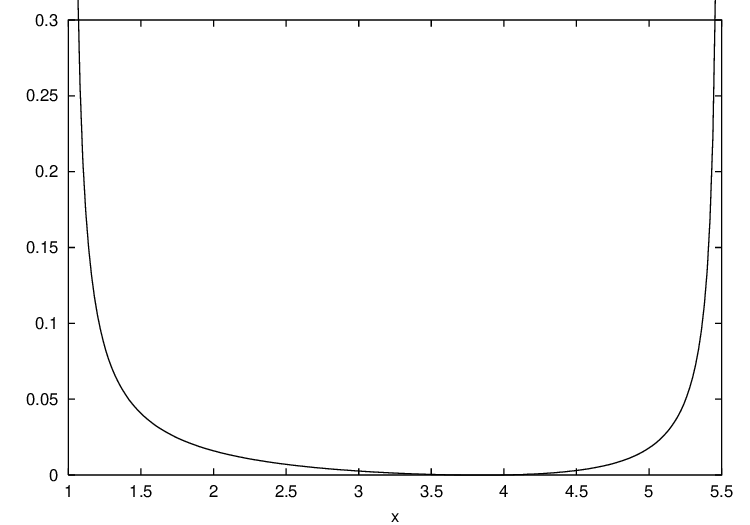}
        \includegraphics[scale=0.5]{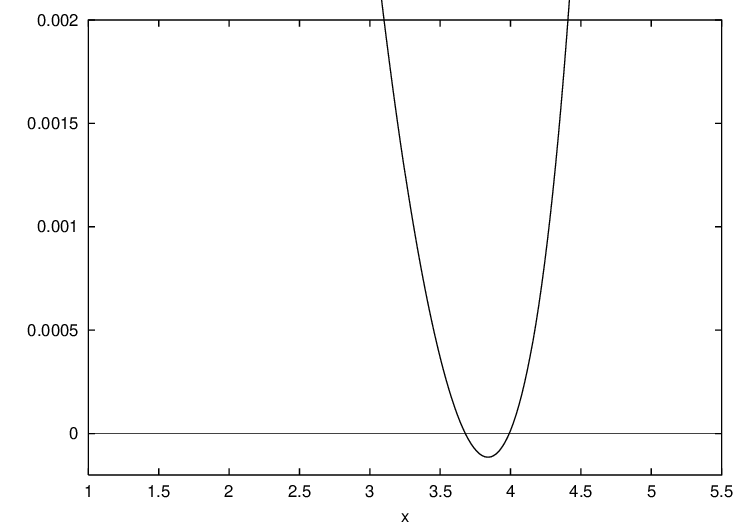}
        \caption{Function $F(x)$. In the right panel the plot is zoomed in order to show its behavior near points corresponding to the Kolmogorov-Zakharov solutions $x=11/3$ and $x=4$. }\label{fig:Vova}
\end{figure}

Important analytical results can be acquired from analysis developed by {\cite{Geogjaev2015}}. Let us estimate collision integral $S_{nl}$ for power-like isotropic functions
\begin{equation}\label{eq:ewx}
    E(\omega)=E_0\left(\frac{\omega}{\omega_0}\right)^{-x}
\end{equation}
One can find an explicit expression
\begin{equation}\label{Snl2omegax}
    S_{nl}^{(\varepsilon)}=8\pi^2 g^{-4} E_0^3 \omega_0^{11} \left(\frac{\omega}{\omega_0}\right)^{11- 3x}{F(x)}
\end{equation}
where function $F(x)$ is calculated by integration in all the  resonant quadruplets \citep[see][for details]{Geogjaev2015}.
Here we used superscript for collision integral in the kinetic equation (\ref{eq:Kinfull}) written for energy. Function $F(x)$ is shown in fig.~\ref{fig:Vova}
This function tends to infinity at $x=1$ and $x=11/2$ when the corresponding integral  diverges. Two zeroes of function $F(x)$ correspond to classic Kolmogorov-Zakharov solutions for direct ($x=4$) and inverse ($x=11/3$) cascades (see eqs. \ref{eq:Kolmogorov_direct}, \ref{eq:Kolmogorov_inverse}).

Similar analytical expression can be obtained for spectral flux of wave energy $P$
\begin{equation}\label{Flux2omegax}
   P=8\pi^2 g^{-4} E_0^3 \omega_0^{12} \left(\frac{\omega}{\omega_0}\right)^{12- 3x} \frac{F(x)}{3x-12}
\end{equation}
For the inverse cascade case $x=11/3$ the energy flux vanishes $F(x)=0$ -- there is no but wave action flux to low frequencies. For $x=4$ -- the direct cascade $F(x)$ tends to zero together with the denominator in (\ref{Flux2omegax}) that gives  finite constant energy flux.

Comparing (\ref{Snl2omegax}) and (\ref{Flux2omegax}) one can get a ratio that can be related with the dissipation coefficient $C_{Ph}$ in (\ref{eq:Sss}) for formally stationary solutions
\begin{equation}\label{Cphil}
    \frac{\omega \partial P/\partial \omega}{P}={3x-12}=C_{Ph}
\end{equation}
For the Phillips spectrum with $x=5$ this ratio gives  $C_{Ph}=3$, i.e. exactly the coefficient for an arbitrary dependence of dissipation function on non-dimensional argument $P\omega^3/g^2$ (see \ref{eq:stat_sol_pow}). Stress again, that this coefficient is fixed for the stationary Phillips spectrum.

\subsection{The Phillips dissipation function for wind sea modelling}
A number of physical constraints should be accounted for constructing a dissipation function for modelling wind sea spectra. First, we express our above findings in terms of conventional dependencies on spectral densities or differential steepness $\mu_w$ rather than on spectral flux as in the primitive model (\ref{eq:Kinext}). Secondly, we introduce a 'switch function' $\Xi$ of a set of physical variables that reflects essential features of strong dissipation associated with wind wave breaking: the breaking usually occurs for sufficiently steep waves or/and for relatively short waves which length is about one order of magnitude shorter than one of dominating (spectral peak) waves \citep{Hwang2007break}. Thus, one can propose the following general form of the dissipation function
\begin{equation}\label{def:DissGen}
   S_{diss}[\omega,E(\omega)]=C_{diss} \mu_w^4 \omega E(\omega)\Xi(\omega/\omega_p,\mu_w,\ldots)
\end{equation}
In this study we use the simplest form of the `switch function' for simulations
\begin{equation}\label{eq:switch}
  \Xi(\omega/\omega_p)=\Theta(\omega/\omega_p-a)
\end{equation}
where $\Theta(\omega/\omega_p-a)$ is the Heaviside function. We let the cut-off parameter to be in the range $a=2\div 4$ \citep{hwang_rough2002,Hwang2007break,Hwang2010} in simulations presented in the next section. Introducing the cut-off parameter $a$ we fix an additional physical scale of the spectral peak frequency $\omega_p$ that makes  wave breakers to be non-uniformly distributed in wave scales that is in contrast with the authentic theory of the \cite{Phillips58} spectrum.

Analysis of the previous section provides an estimate of the dissipation coefficient $C_{diss}$   for the Phillips spectrum (\ref{eq:ewx}). Assuming $a=0$ (no breaking cut-off) one has  the balance of dissipation and nonlinearity at
\begin{equation}\label{CPhillips}
    C_{diss}=C_{Phillips} = 8\upi^2 F(5)\approx 2.03
\end{equation}
With the  approximate expression for $F(x) $ at $x\to 11/2$ one gets {\cite[][]{Geogjaev2015}}
\[
F(x)=\frac{0.0129}{11/2-x}, \qquad x\to 11/2
\]
The theoretical estimate of the Phillips dissipation coefficient $C_{Phillips}$ (\ref{CPhillips}) gives a good reference in our numerical study for the dissipation coefficient $C_{diss}$ in (\ref{def:DissGen}).

\section{Numerical results}
Numerical simulation of the kinetic equation (\ref{eq:Kinfull}) has been carried out in order to check the simple theory presented above. The WRT algorithm \cite[][]{Webb78,Tracy82} was used for calculation collision integral. The dissipation function was calculated in accordance with definition (\ref{def:DissGen}) with the cut-off function (\ref{eq:switch}), i.e. in terms of `observable' parameters of wave spectra. In this paper we present the very first results of simulations within the simplest setup. We studied duration-limited evolution (spatially homogeneous) of initially isotropic spectrum. The initial spectrum is given by JONSWAP-like formula for frequency spectrum of energy
\begin{equation}
\label{eq:JONSWAP}
E(\omega)= \frac{\alpha g^2}{ \omega_p}\omega^{-4} \exp \left(-\frac{5}{4}\left(\frac{\omega}{\omega_p}\right)^{-4}\right) \gamma^G
\end{equation}
where
\begin{equation}
G=\exp\left(-\frac{(\omega-\omega_p)^2}{2\sigma_p^2\omega_p^2}\right)
\end{equation}
with default peakedness parameters $\gamma=3.3,\, \sigma_p=0.08$. The initial wavelength $\lambda_p=2\pi g/\omega_p^2$ corresponds to rather long swell $\lambda_p\approx 240$ m. Minimal frequency in simulations $f_{min}=0.02$Hz, maximal -- $f_{max}=2$Hz, peak frequency $f_{p}=0.079$Hz, the dissipation cutoff is set at four times higher frequency $f_{diss}=0.319$Hz. With logarithmic grid  each frequency domain (low bound $[f_{min}-f_p]$, $[f_p-f_{diss}]$) contains approximately the same number of grid points. In absence of wave input  slow spectral downshift does not change essentially this division.

\begin{figure}
\begin{center}
\includegraphics[scale=0.43]{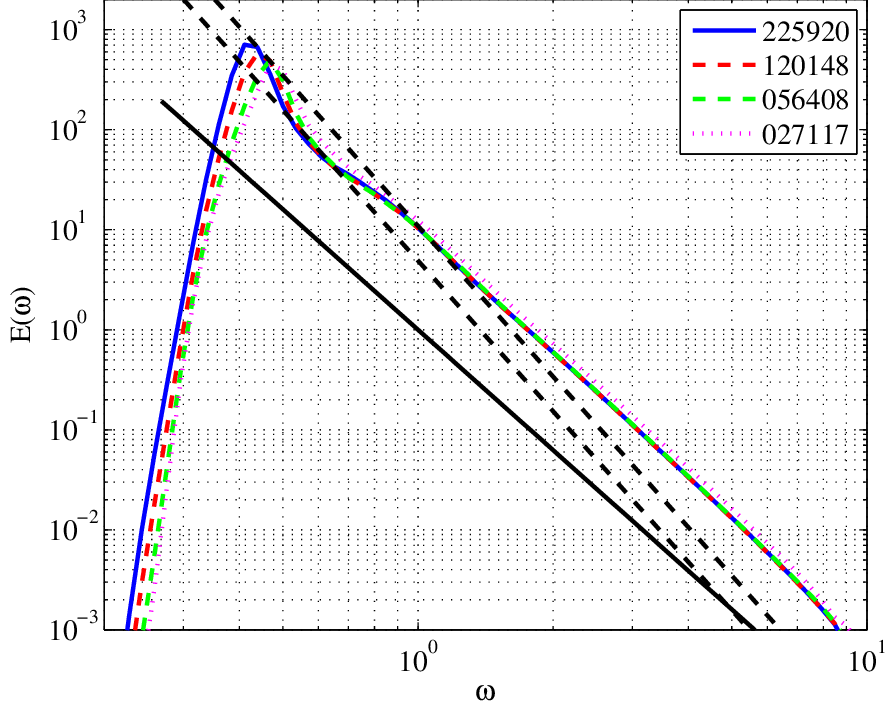}
\includegraphics[scale=0.43]{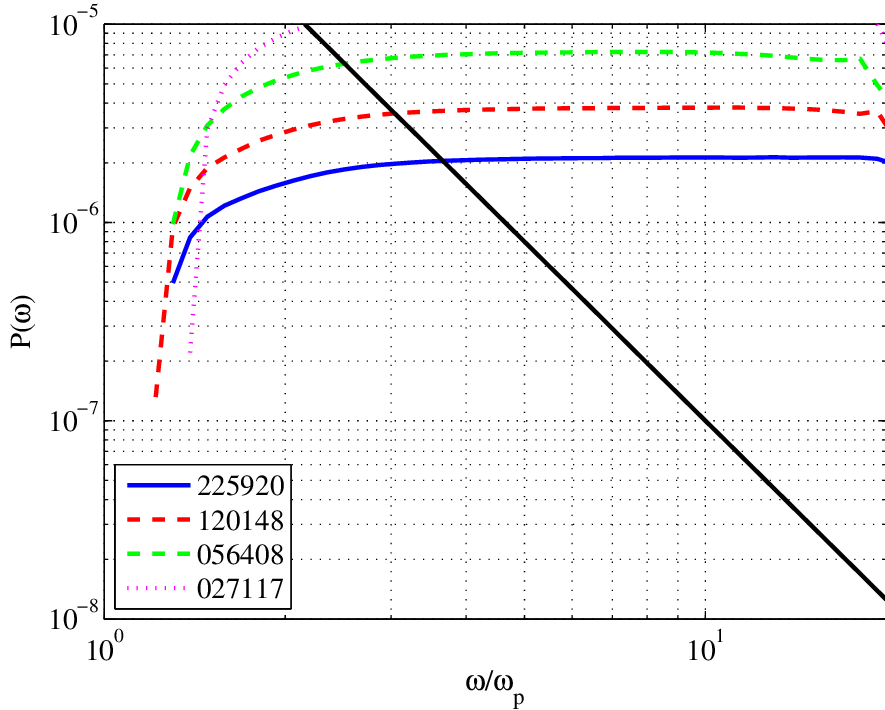}\\
\includegraphics[scale=0.43]{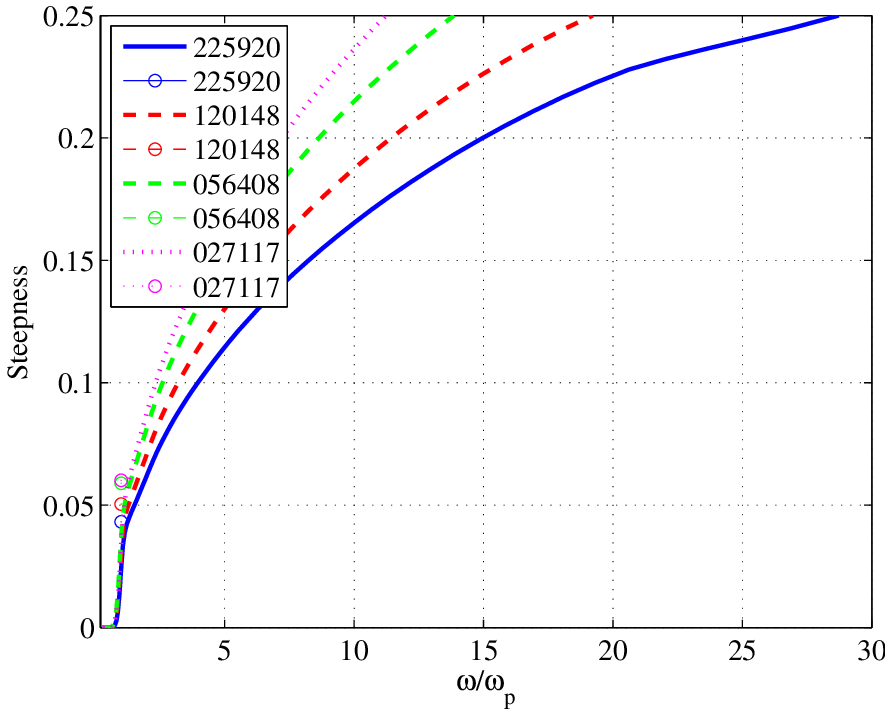}
\includegraphics[scale=0.43]{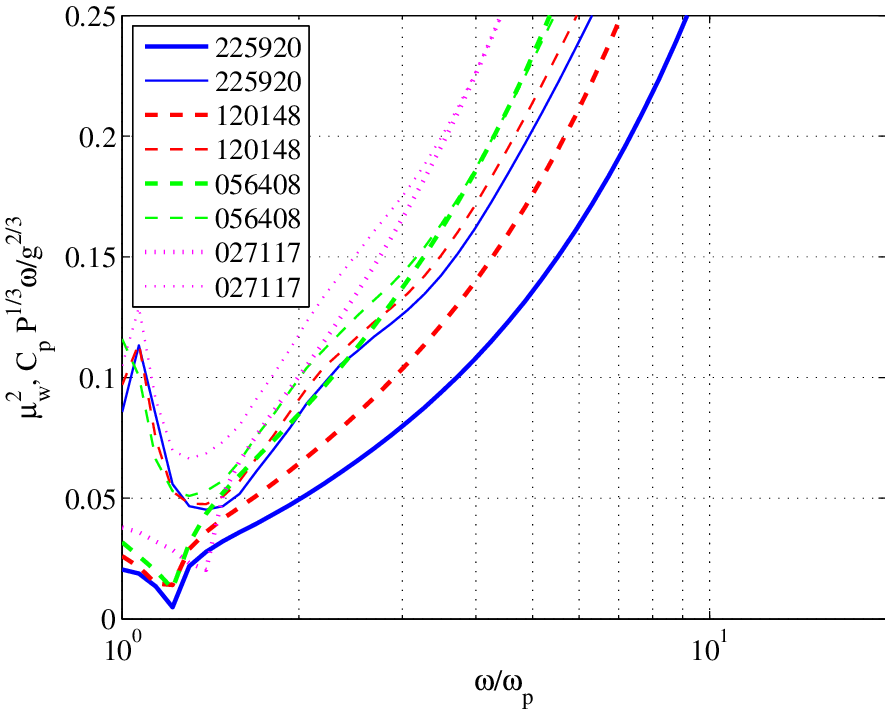}
 \end{center}
  \caption{The kinetic equation solutions for reference case of swell: top-left  -- frequency spectra of energy, hard straight line -- $E(\omega)\sim \omega^{-4}$, dashed -- $\omega^{-5}$; top-right  -- energy flux (curves), straight line corresponds to law $P\sim \omega^{-3}$; bottom-left -- cumulative (lines) and integral wave steepness (symbols at $\omega/\omega_p=1$); bottom-right -- differential steepness   defined in terms of spectral flux (bold line) and parameter $\mu_w$. Time in seconds is given in legend.}\label{fig:swell0}
\end{figure}

The reference case of `pure swell' is presented in fig.~\ref{fig:swell0}. As it was demonstrated in \cite{BPRZ2005} the solution evolves to a self-similar shape. In our case we see a transition to a self-similar behavior rather than the self-similarity itself because of relatively short time of evolution and realistic low wave steepness. Energy spectra (top fig.~\ref{fig:swell0}) show downshifting and a peakedness growth. The energy spectral flux tends to constant magnitudes in a wide range of non-dimensional frequencies ($3 < \omega/\omega_p <20$). Integral wave steepness $\mu_p$ (circles in bottom fig.~\ref{fig:swell0}) is going down with time as well as curves of cumulative steepness $\mu_c$ which is growing function of frequency. The differential steepness shows similar behavior: steep growth with frequency and decay with time.

\begin{figure}
\begin{center}
\includegraphics[scale=0.43]{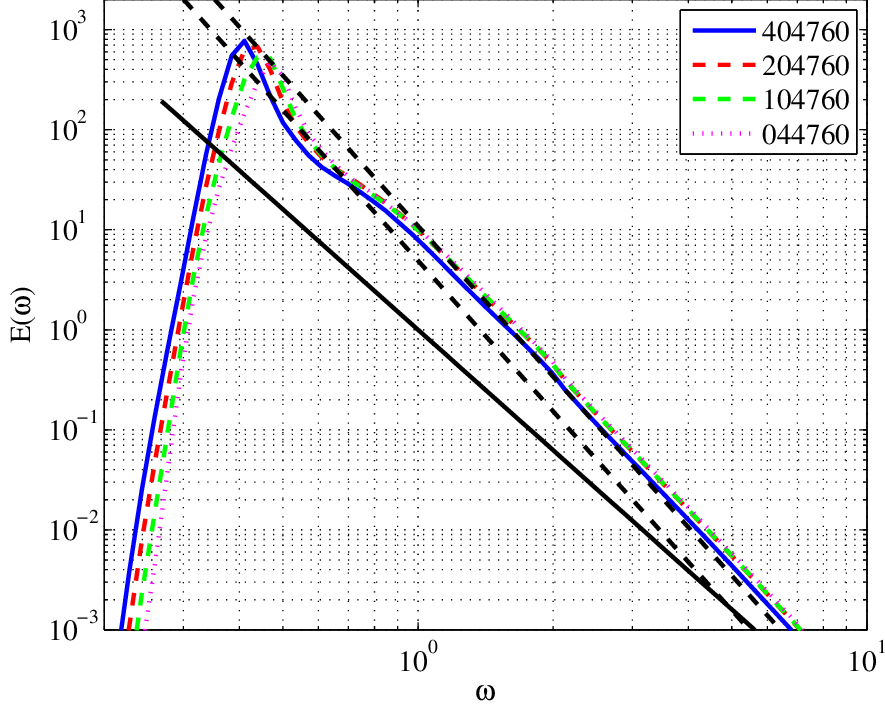}
\includegraphics[scale=0.43]{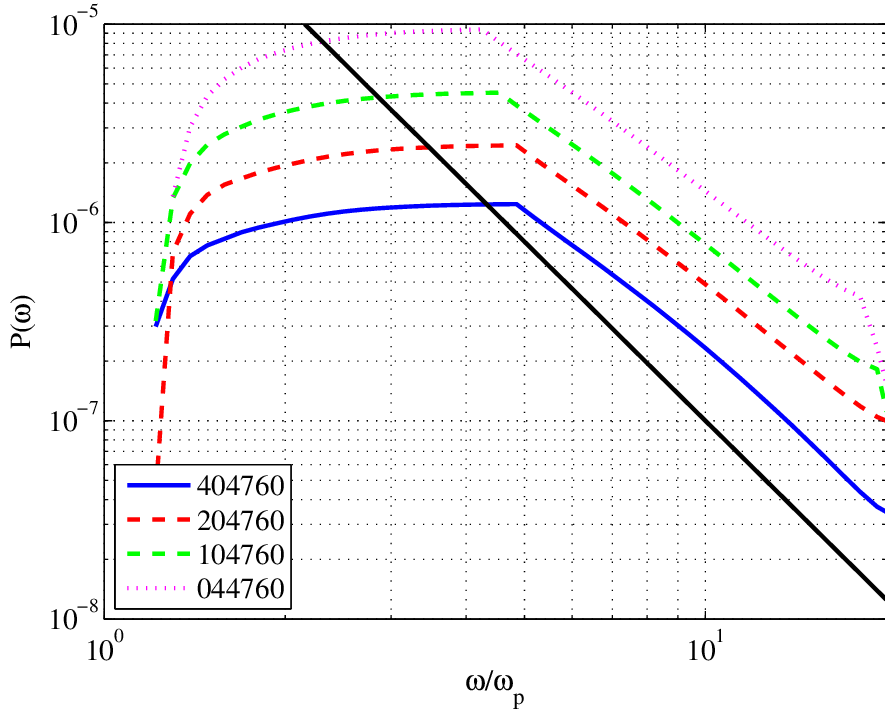}\\
\includegraphics[scale=0.43]{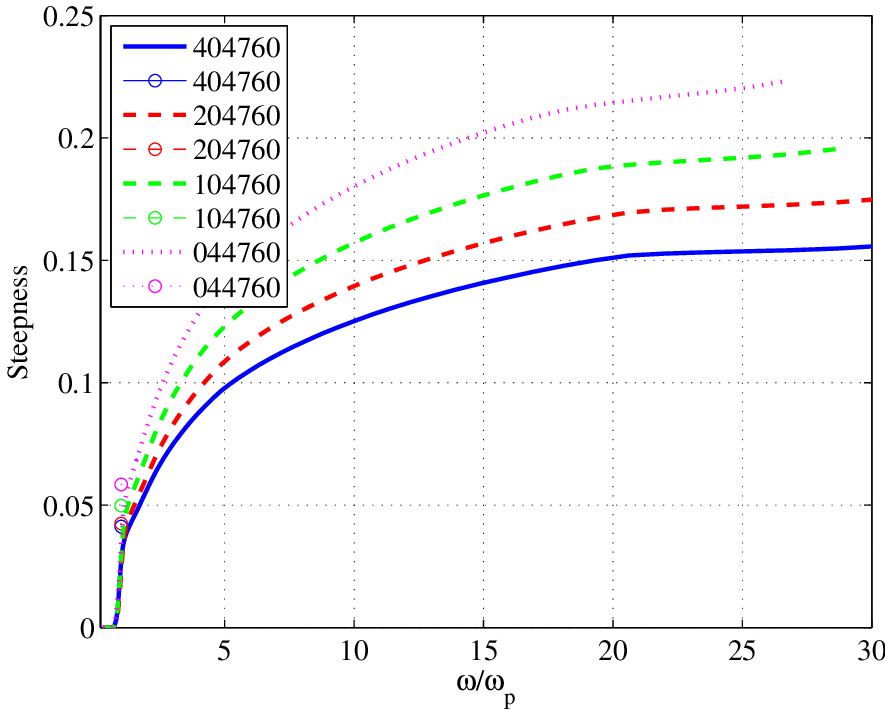}
\includegraphics[scale=0.43]{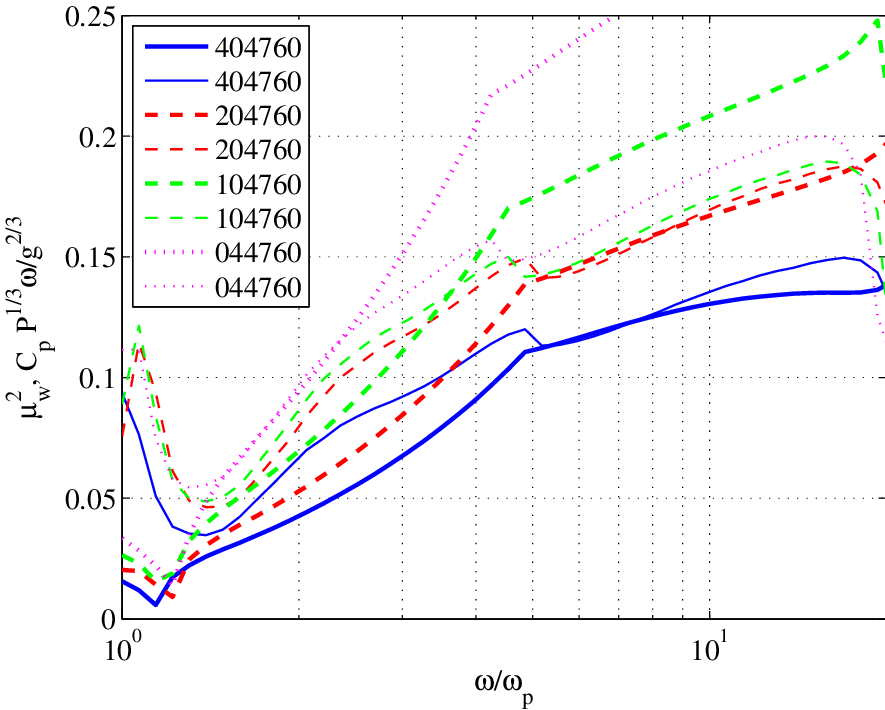}
 \end{center}
  \caption{The kinetic equation solutions for dissipation function (\ref{def:DissGen}) with $C_{Ph}=1.22$ and dissipation cutoff $\omega_{diss}=2$rad$^{-1}$: top-left  -- frequency spectra of energy, hard straight line -- $E(\omega)\sim \omega^{-4}$, dashed -- $\omega^{-5}$; top-right  -- energy flux (curves), straight line corresponds to law $P\sim \omega^{-3}$; bottom-left -- cumulative (lines) and integral wave steepness (symbols at $\omega/\omega_p=1$); bottom-right -- differential steepness   defined in terms of spectral flux (bold line) and parameter $\mu_w$. Time in seconds is given in legend.}\label{fig:sw1_3}
\end{figure}

Fig.~\ref{fig:sw1_3} presents results for dissipation function with coefficient $C_{diss}=1.22=0.6 C_{Phillips}$ and dissipation cutoff frequency $\omega_{diss}=2$rad$^{-1}$ . This coefficient is almost two times less than our estimate of the Phillips coefficient $C_{Phillips}=2.03$ defined for the Phillips exponent $(-5)$. One can see appearance of essentially steeper spectrum tail as compared to `pure swell' of fig.~\ref{fig:swell0} and dramatic reduction of all the steepness parameters $\mu_p,\, \mu_w$ while for cumulative steepness $\mu_{c}$ one can see a sort of saturation. At the same time, the differential steepness $\mu_w$ shows a gradual growth with frequency. In bottom-right fig.~\ref{fig:sw1_3} we give curves in accordance with two possible definitions of dissipation, first, in terms of spectral flux (\ref{eq:New_diss}), second, as a function of wave steepness $\mu_w$ (\ref{def:DissGen},\ref{eq:switch}) and conventional presentation of dissipation in wave models (\ref{eq:diss_PhillipsNk}, \ref{eq:diss_PhillipsNw}).

Higher dissipation rate $C_{Ph}=1.9$ approximately $ 5$\%  lower than the Phillips dissipation coefficient  $C_{Phillips}=2.03$  shows very strong tendency to Phillips spectrum $E(\omega)\sim \omega^{-5}$. The bottom-right panel fixes two features: first, the differential steepness is tending to saturation, second, two alternative definitions of the parameter, in terms of spectral flux (bold lines) and as function of wave amplitude (thin line), appear to be close to each other.
\begin{figure}
\begin{center}
\includegraphics[scale=0.43]{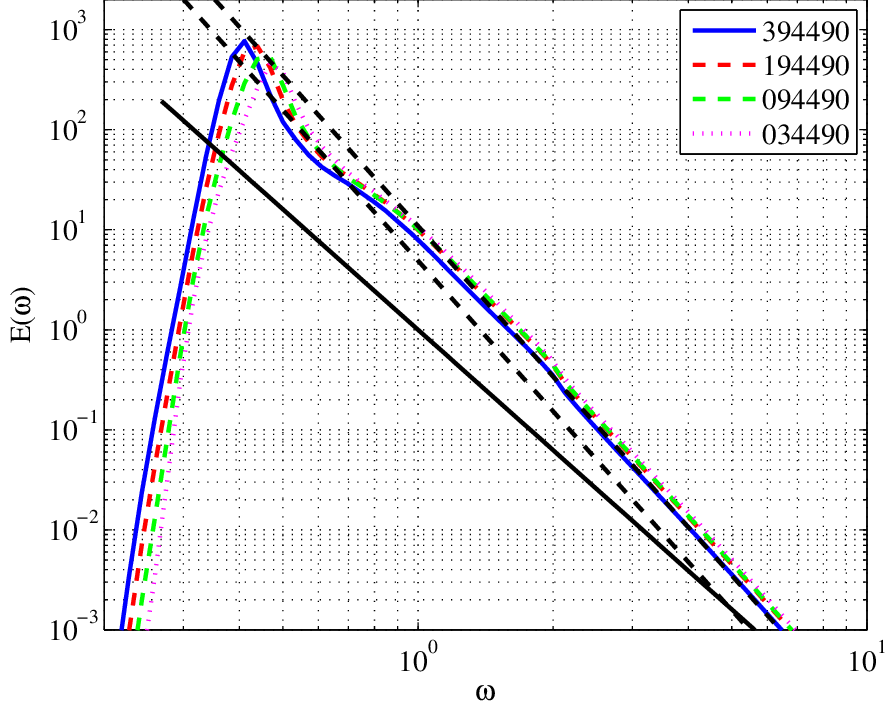}
\includegraphics[scale=0.43]{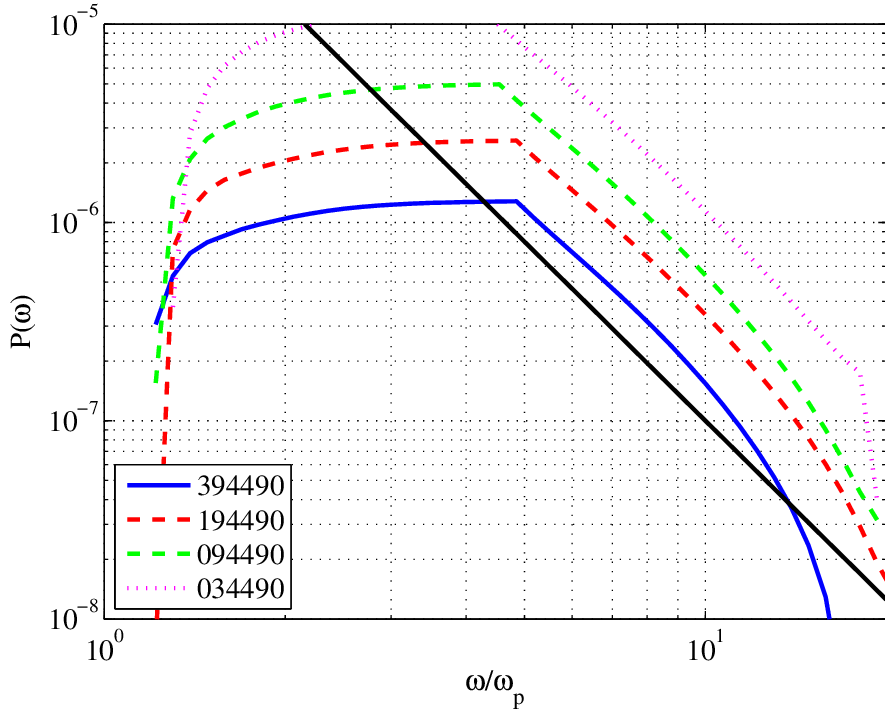}\\
\includegraphics[scale=0.43]{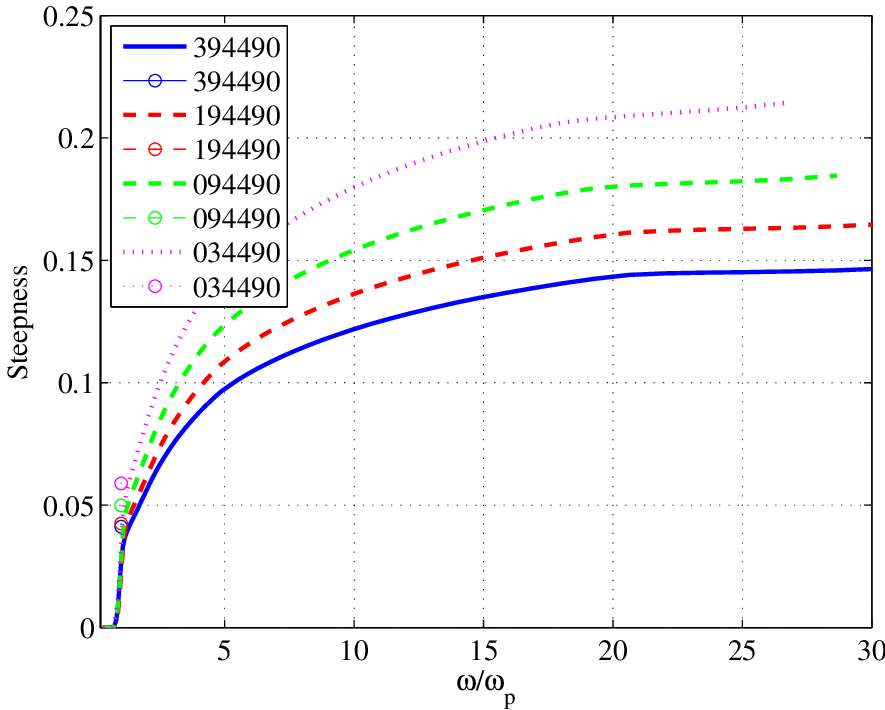}
\includegraphics[scale=0.43]{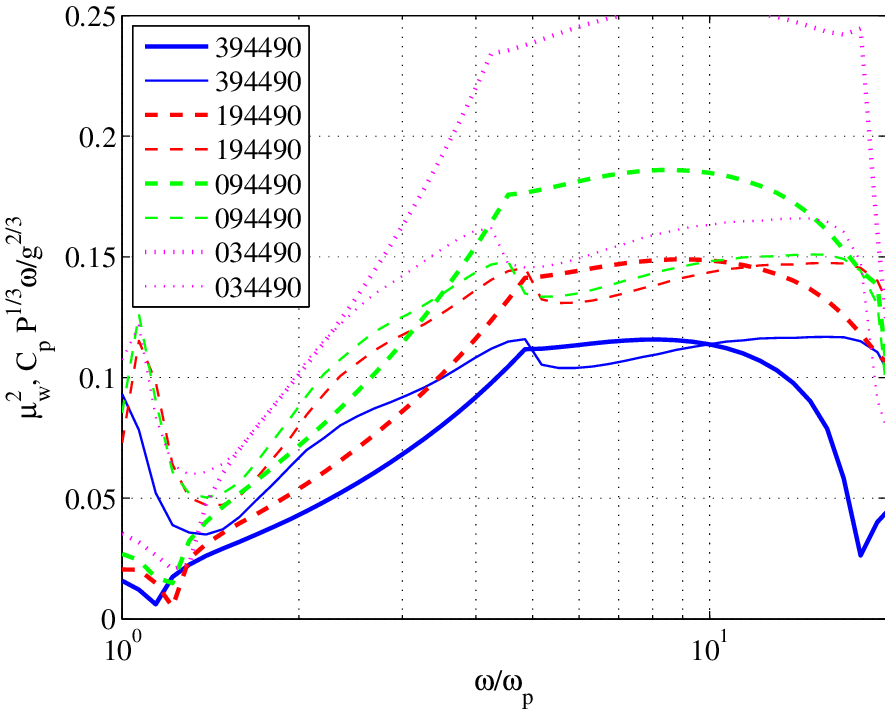}
 \end{center}
  \caption{The kinetic equation solutions for dissipation function (\ref{def:DissGen}) with $C_{Ph}=1.9$ and dissipation cutoff $\omega_{diss}=2$rad$^{-1}$: top-left  -- frequency spectra of energy, hard straight line -- $E(\omega)\sim \omega^{-4}$, dashed -- $\omega^{-5}$; top-right  -- energy flux (curves), straight line corresponds to law $P\sim \omega^{-3}$; bottom-left -- cumulative (lines) and integral wave steepness (symbols at $\omega/\omega_p=1$); bottom-right -- differential steepness   defined in terms of spectral flux (bold line) and parameter $\mu_w$. Time in seconds is given in legend.}\label{fig:sw125_5}
\end{figure}

\begin{figure}
\begin{center}
\includegraphics[scale=0.43]{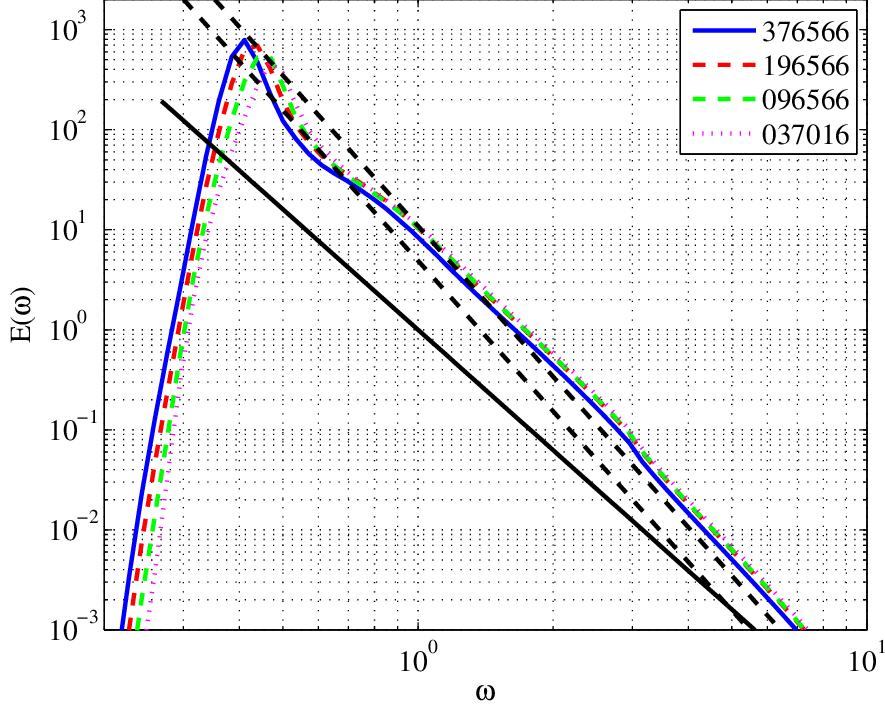}
\includegraphics[scale=0.43]{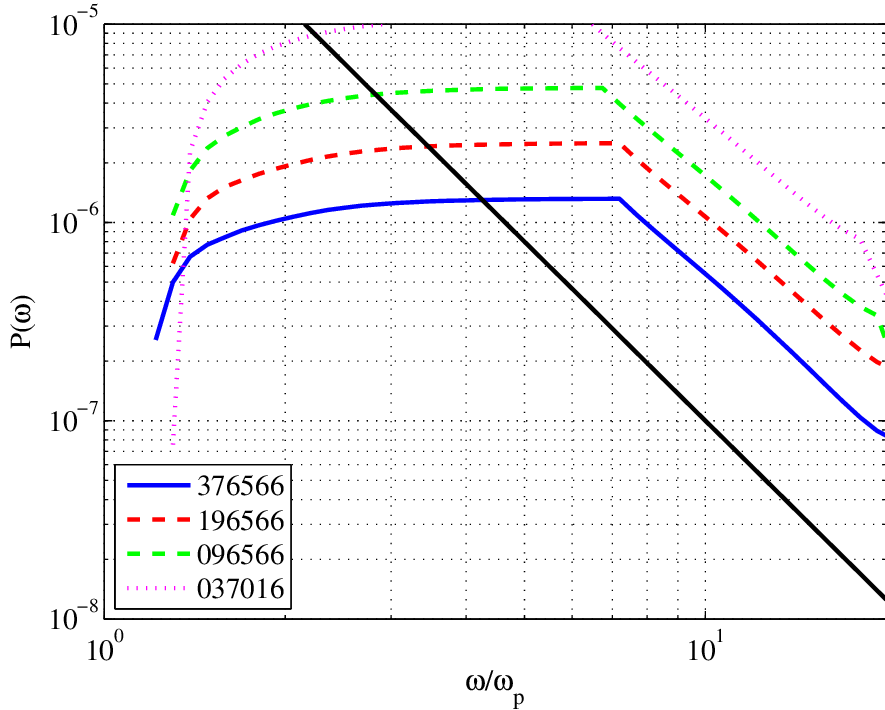}\\
\includegraphics[scale=0.43]{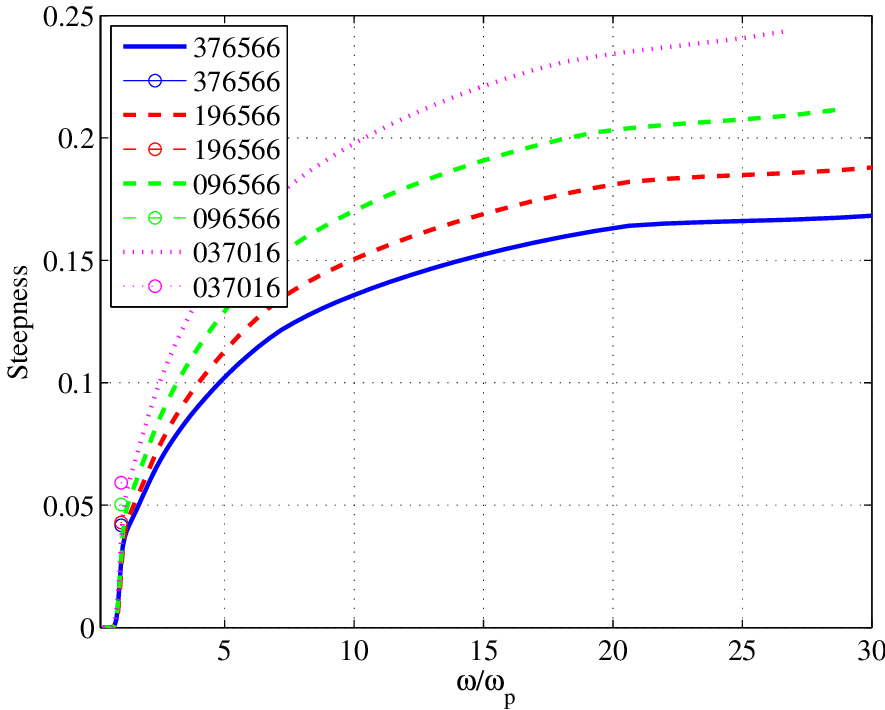}
\includegraphics[scale=0.43]{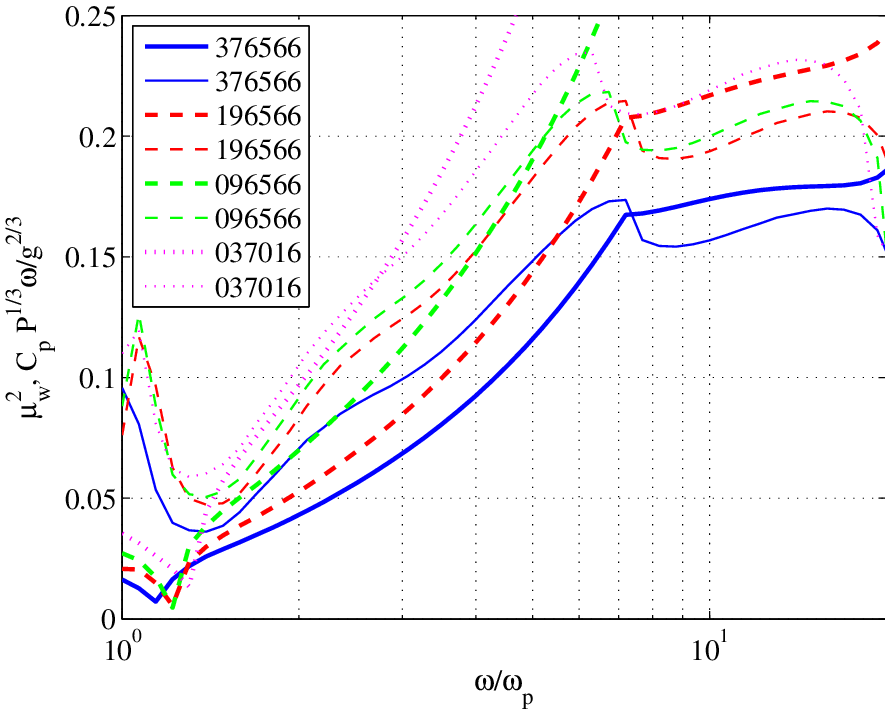}
 \end{center}
  \caption{The kinetic equation solutions for dissipation function (\ref{def:DissGen}) with $C_{Ph}=1.9$ and dissipation cutoff $\omega_{diss}=3$rad$^{-1}$: top-left  -- frequency spectra of energy, hard straight line -- $E(\omega)\sim \omega^{-4}$, dashed -- $\omega^{-5}$; top-right  -- energy flux (curves), straight line corresponds to law $P\sim \omega^{-3}$; bottom-left -- cumulative (lines) and integral wave steepness (symbols at $\omega/\omega_p=1$); bottom-right -- differential steepness   defined in terms of spectral flux (bold line) and parameter $\mu_w$. Time in seconds is given in legend.}\label{fig:sw125_5_3}
\end{figure}

One more example illustrates the effect of dissipation cutoff. In fig.~\ref{fig:sw125_5_3} all the parameters of simulation are the same as in previous case but dissipation cutoff is set at $\omega_{diss}=3$rad$^{-1}$. One can see the same effect: very strong tendency of the solution to Phillips' asymptotics and saturation of all the steepness parameters ($\mu_p,\,\mu_c,\,\mu_w$)

\section{\label{discussion}Conclusions and Discussion}
In this work we proposed to reanimate the Phillips idea of wave spectra as a result of a balance of different physical mechanisms. We combine effects of nonlinear transfer due to four-wave resonant interactions of weakly nonlinear deep water waves and  dissipation dealing with wave breaking. These two effects can be co-existing in quite good symbiosis for the famous Phillips spectrum $E(\omega)\sim \omega^{-5}$ (\ref{eq:Phillips_spec}): just this spectral shape makes asymptotic kinetic equation to be formally valid in the whole range of wave scales. Thus, we have a good example of the generalized Phillips spectrum in this case.

There is one more point of interest in competition  of dissipation and nonlinearity for the Phillips spectrum. Stationarity of this spectrum requires certain magnitude of dissipation coefficient -- the Phillips coefficient. This is in contrast to widely accepted models of wave dissipation where dissipation function is assumed to be arbitrary: the spectral shaping does not depend on the dissipation magnitude. We gave an analytical estimate of the Phillips dissipation coefficient $C_{Phillips}\approx 2.03$.

Our simulation within the simplest setup of slowly evolving isotropic wave field justified our theoretical results. Shape of wave spectra are shown to be sensitive to magnitude of wave dissipation. For values close to the found theoretical value $C_{Phillips}\approx 2.03$ the spectra are tending to the Phillips classic tail $\omega^{-5}$ and appeared to be long-lived. Deviation from this value gives alternative spectral slopes. Variety of spectral slope is well-known from experimental studies. Collection of spectral slopes derived from buoy data have been presented, say, by  \cite{Liu1988,Liu1989}. Fig.~8 by \cite{Liu1988} shows a wide range of the exponents: $90$\% of them fall into range $3-7$ with a maximum between $4$ and $5$. This high dispersion  has been attributed to  `an equilibrium range' of wave spectra with no clear idea what mechanisms are responsible for this equilibrium and why the Phillips spectrum $\omega^{-5}$ is emphasized in this experimental collection.

Our very first numerical experiments with the new dissipation function  showed robustness of the effect of nonlinear dissipation. Its effect leads to saturation of wave steepness both in integral ($\mu_p$) or differential ($\mu_w$) quantities. This feature is of great importance for further simulation of wave spectra with both effects of wave input and dissipation taken into account. High nonlinearity of the new dissipation function suppresses effect of wave pumping in high frequency completely and, thus, the classic Phillips spectrum continues to play an important role in general case.

\vskip 0.5cm
Authors are thankful to Vladimir Geogjaev for providing fig.1 and helpful discussion of results. The research  was conducted under   grant of Russian Science Foundation N14-22-00174.  This support is gratefully acknowledged.




\end{document}